\documentclass{bvm} 

\makeatletter
\let\blx@rerun@biber\relax
\makeatother

\begin{document}

\newcommand{\bvmyear}{2023}

\selectlanguage{english} 


\title{Multi-task Learning To Improve Semantic Segmentation Of CBCT Scans Using Image Reconstruction\footnote{Accepted at German Conference on
Medical Image Computing 2024}}


\titlerunning{Multi-task Learning to Improve Segmentation}

\author{
    Maximilian~E. \lname{Tschuchnig} \inst{1,3},
    Julia \lname{Coste-Marin} \inst{2},
    Philipp \lname{Steininger} \inst{2},
    Michael \lname{Gadermayr} \inst{1} 
}


\authorrunning{Tschuchnig et al.}

\institute{
\inst{1} Salzburg University of Applied Sciences\\
\inst{2} MedPhoton GmbH\\
\inst{3} University of Salzburg\\
}

\email{maximilian.tschuchnig@fh-salzburg.ac.at}

\maketitle

\begin{abstract}
Semantic segmentation is a crucial task in medical image processing, essential for segmenting organs or lesions such as tumors.
In this study we aim to improve automated segmentation in CBCTs through multi-task learning. To evaluate effects on different volume qualities, a CBCT dataset is synthesised from the CT Liver Tumor Segmentation Benchmark (LiTS) dataset. To improve segmentation, two approaches are investigated. First, we perform multi-task learning to add morphology based regularization through a volume reconstruction task. Second, we use this reconstruction task to reconstruct the best quality CBCT (most similar to the original CT), facilitating denoising effects. We explore both holistic and patch-based approaches. 
Our findings reveal that, especially using a patch-based approach, multi-task learning improves segmentation in most cases and that these results can further be improved by our denoising approach. 
\end{abstract}

\section{Introduction}
Segmentation is a crucial field of automated medical image analysis, for a multitude of applications, from segmenting low-level (e.g., nuclei) to high-level tissue structure (differentiation between cancer, stroma, and necrosis). 
Classical segmentation methods (thresholding, region-based, watershed), though useful for both 2D and 3D segmentation, often struggle with the complexity of pathological and anatomical structures, particularly high-level structures observed in medical imaging. Deep learning methods exhibit the state-of-the-art for semantic segmentation, commonly employing encoder-decoder models~\cite{isensee2019nnu,hatamizadeh2022unetr,xie2021segformer}. These models are trained to encode latent information in a bottleneck and then use this information to reconstruct a mask (segmentation) or the original volume (image reconstruction).
Fully-convolutional networks, exemplified by Unet~\cite{ronneberger2015u}, utilize this encoder-decoder structure to generate segmentation masks. Models like nn-unet~\cite{isensee2019nnu} extend Unet's architecture to 3D and demonstrate state-of-the-art performance on various computed tomography (CT) segmentation tasks~\cite{bilic2023liver}. 
UnetR~\cite{hatamizadeh2022unetr} and SegFormer~\cite{xie2021segformer} adapt nn-unet by incorporating transformer blocks like multihead attention.

Multi-task learning aims to share knowledge over multiple tasks of a common model for knowledge transfer, regularization and data efficiency. E.g. multi-task learning approaches have previously been investigated by Weninger et al.~\cite{weninger2020multi} to reduce the amount of needed labelled training data by adding reconstruction to a segmentation task. Similarly, Mlynarski et al.~\cite{mlynarski2019deep} include a task of 2D tumor detection score estimation. Both approaches use encoder-decoder structures. Therefore, the addition of multiple tasks increase the amount of data that can be used to train their encoder. 

We propose a novel, model agnostic, extension to semantic segmentation. By formulating model optimisation through multi-task learning of segmentation and image reconstruction we facilitate morphology based regularisation. For experimental validation, we select the state-of-the-art for automated semantic segmentation, 3D nn-unet~\cite{isensee2019nnu,bilic2023liver}. The evaluation is conducted using the well-established Liver Tumor Segmentation Benchmark (LiTS)~\cite{bilic2023liver}. To assess our proposed adaptation at different data quality, we synthesise cone-beam CT (CBCT) from the original LiTS, with varying numbers of used projections employed for CT reconstruction. 
We further evaluate the effect of using the CBCT corresponding to the highest quality volumes as the reconstruction target, facilitating a denoising effect. 

\section{Method}
\begin{figure*}[tb]
  \centering
  \includegraphics[width=1\textwidth]{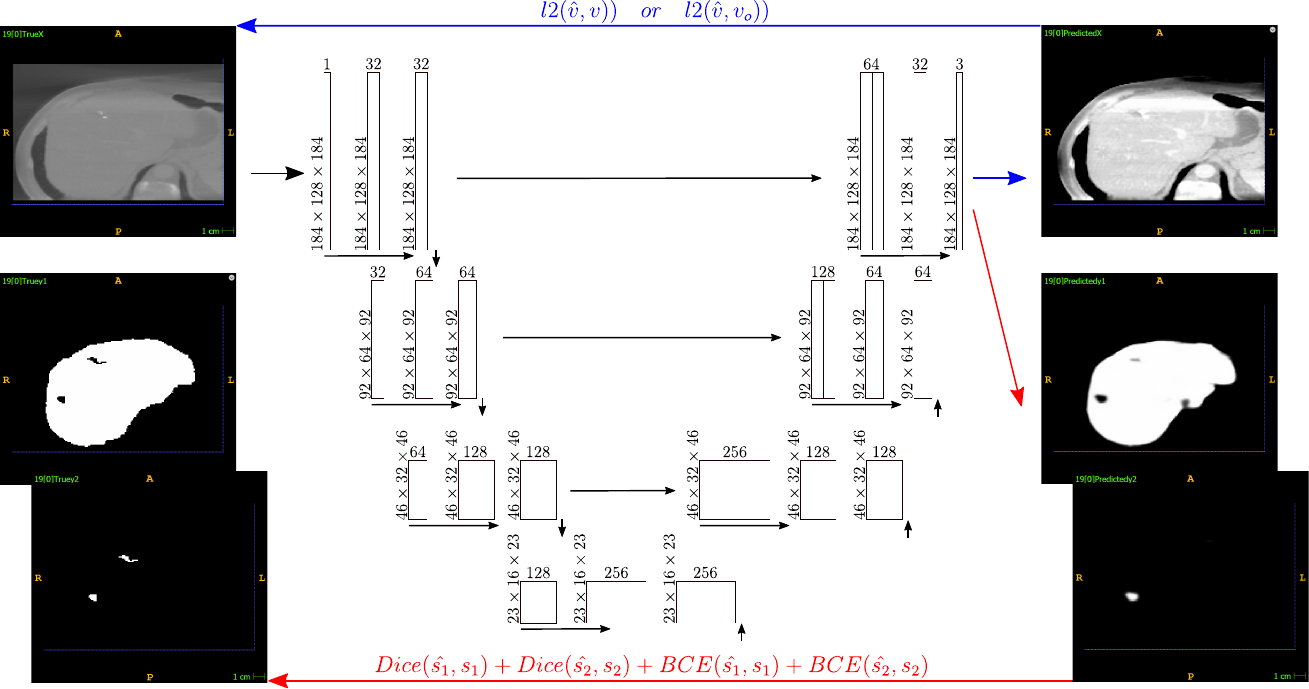}
  \caption{Segmentation unet (red path) and the reconstruction task (blue), facilitating multi-task learning. The volumes are used in patches and holistically (downscaled by $2$). The multi-task trained model performs segmentation and reconstruction at once (red and blue paths). The final loss includes Dice, BCE and L2 to facilitate training both segmentation and reconstruction.}
  \label{fig:proposedUnet}
\end{figure*}

Our model agnostic approach, shown applied to a 3D nn-unet in Fig.~\ref{fig:proposedUnet}, is to 1) add a regularisation term to semantic segmentation through multi-task learning with the additional task of image reconstruction and 2) add denoising effects to the reconstructions of 1) by always reconstructing the highest quality volume.

In detail, the proposed multi-task method describes a generic adaptation of the training stage of semantic segmentation approaches. Additionally to performing segmentation the multi-task model mt-c is trained to reconstruct the current volume for regularisation. Assuming binary cross-entropy (BCE) as the segmentation and L2 as the reconstruction loss, a combined loss can be formulated as $Loss = BCE(\hat{s}, s) + l2(\hat{v}, v)$. With $s$ as the real and $\hat{s}$ as the predicted segmentation as well as $v$ denoting the current and $\hat{v}$ the reconstructed image. The intuition is to explicitly regularise the learned segmentation using morphology. By synthesizing CBCTs from LiTS using different amounts of used projections, different quality CBCT volumes are generated. By replacing $v$ with the highest quality volume $v_o$, the model is further trained to denoise lower visual quality volumes. We refer to this method as mt-b (best quality).

As shown in Fig.~\ref{fig:proposedUnet}, a 3d nn-unet~\cite{ronneberger2015u,isensee2019nnu} is chosen as the baseline (red path). This unet consists of an encoder with $3$ double convolution layers and $3 \times 3 \times 3$ convolutional kernels, connected by 3d max pooling. The latent space consists of one double convolution block leading into the unet decoder which mirrors the encoder.
As is typicall for unet, each double convolutional output in the encoder is also connected to the decoder double convolutional block of the same order. Additionally, one 3d convolutional layer is added to the decoder with a filter size of $1 \times 1 \times 1$ and the number of filters set to the amount of segmentation classes. The rest of the convolutional layers are set to $\{32, 64, 128, 256\}$ filters along the stream direction in the encoder. Batch norm is applied after each layer in the double convolutional blocks. The model is trained utilizing a sum ($Loss_1$) of BCE and Dice $Loss_1 = BCE(\hat{s}, s) + Dice(\hat{s}, s)$.

We apply the multi-task adaptation, shown in Fig.~\ref{fig:proposedUnet} (combining both the red and blue path), on the introduced baseline. The last double convolutional layer output is used multistream, with one stream feeding into the baseline segmentation layer and the other stream feeding into a 3d convolutional layer with filter size $1 \times 1 \times 1$ and a linear activation function to facilitate reconstruction. The baseline loss is adapted by adding the l2 $Loss_2 = \alpha \cdot (BCE(\hat{s}, s) + Dice(\hat{s}, s)) + (1 - \alpha) \cdot l_2(\hat{v} - v)$ with $\alpha=0.8$. Depending on the setup, the optimal $v_o$ is used instead of $v$ in training.

\subsection{Dataset}
Evaluation is performed using the LiTS dataset. LiTS consists of $131$ abdominal CT scans in the training set and $70$ test volumes. The $131$ train volumes include segmentations of 1) the liver and 2) liver tumors. The dataset contains data from $7$ different institutions with a diverse set of liver tumor diseases. 
For further information we refer to Bilic et al.~\cite{bilic2023liver}. To evaluate effects on vastly different volume qualities, the LiTS dataset was converted into synthetic CBCT scans. Corresponding projections were simulated per CT volume, with the number of projections $n_p \in \{490,256,128,64,32\}$. 
$490$ projections was chosen as the default configuration to show a visual quality similar to the original CT 
and $32$ as the lowest configuration, showing little information and many artifacts.

\subsection{Experimental Setup}
We evaluate the introduced methods holistically and patch based on liver and liver tumor segmentation~\cite{bilic2023liver,araujo2022liver,han2022effective,wang2022automatic}. The segmentations are evaluated using the Dice score. A threshold is applied to the model output before Dice calculation to establish binary labels. This threshold is set to $0.5$ in all models and setups. To fit the holistic data on an NVIDIA RTX A6000 grafics card the volumes are downscaled by the factor $2$. For patched segmentation patches of the size $192 \times 192 \times 192$ are extracted from the full-scale CBCT data. On inference, the patch segmentations are re-agregated before calculating a holistic Dice. 

Models and setups are evaluated using the different quality CBCT datasets with the quality depending on the amount of used projections for simulation $n_p \in \{490,256,128,64,32\}$. A further investigated parameter is the reconstruction target with the two investigate setups, mt-c and mt-b. The setup mc-c uses the input volume of the current quality as its reconstruction target while mt-b always uses the best possible quality input volume ($v_o$) as its reconstruction target. Therefore, in the case $n_p=490$ both mt-b and mt-c lead to the same results. All setups are trained and evaluated $4$ times to facilitate stable results with the same random splits for results comparability. Since annotations are not available for the LiTS test dataset, the test dataset was disregarded for this publication and the original train set was separated into training-validation-testing data~\cite{araujo2022liver,han2022effective}. The separation was performed using the ratio 0.7, 0.2 and 0.1 ($training:0.7$, $validation:0.2$, $testing:0.1$).

\section{Results}

\begin{figure}[ht]
    \centering
    \includegraphics[width=0.32\textwidth]{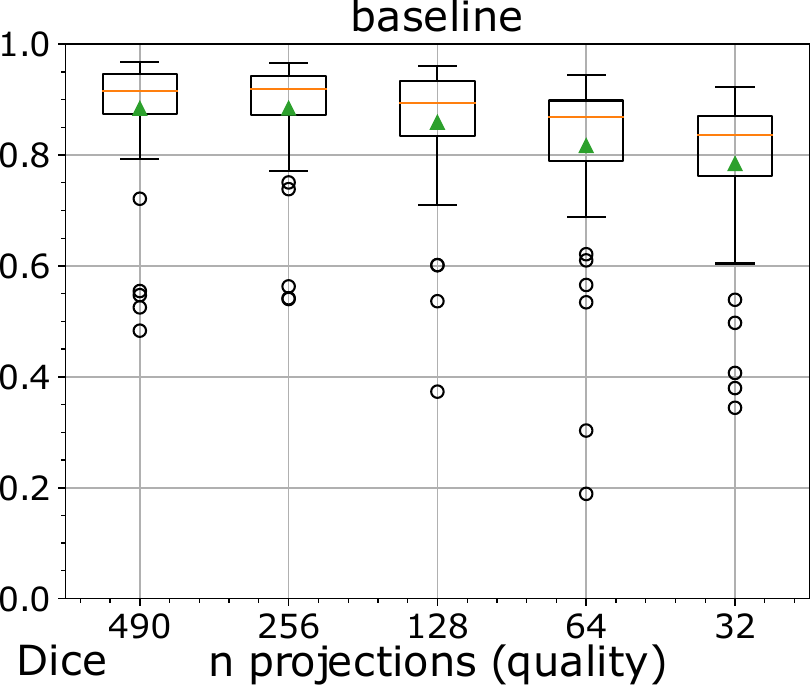}
    \includegraphics[width=0.32\textwidth]{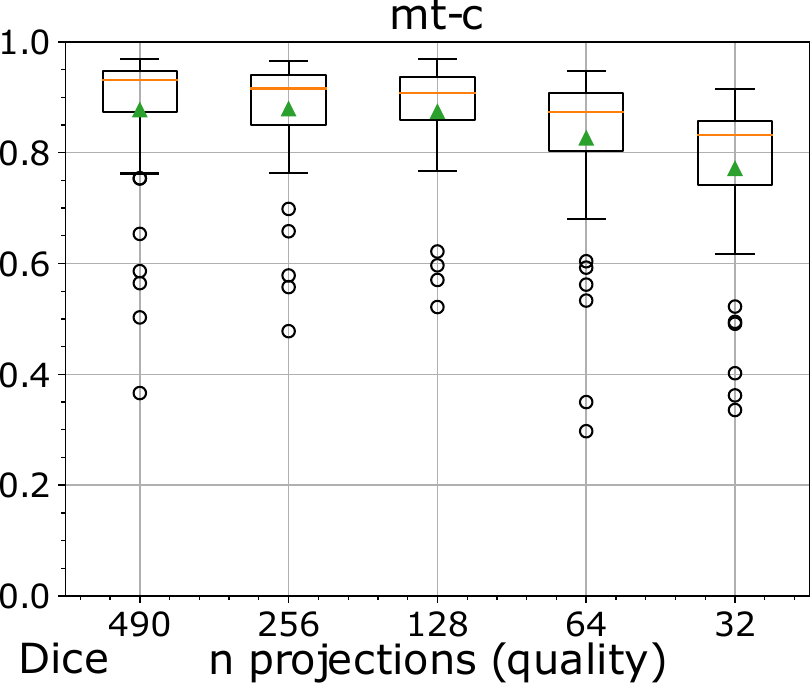}
    \includegraphics[width=0.32\textwidth]{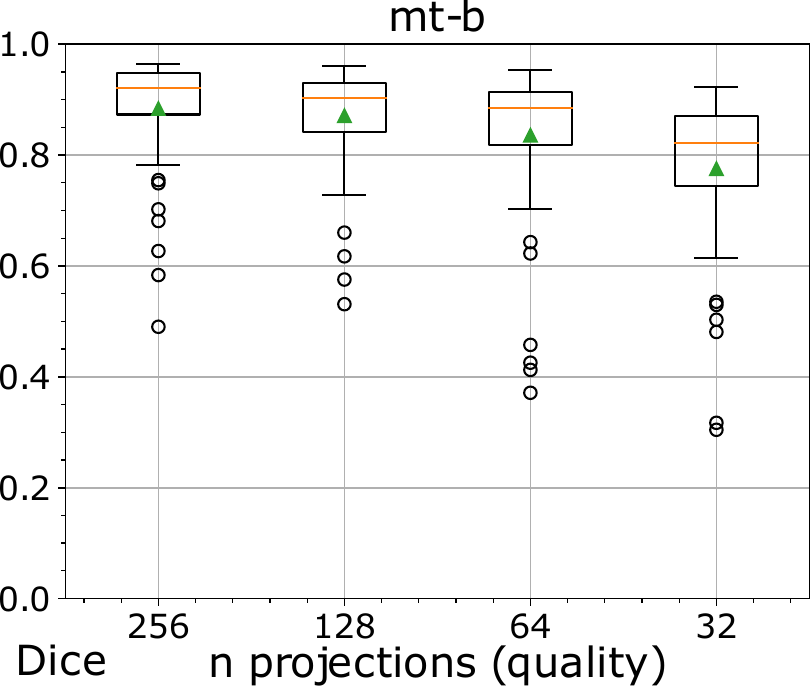} \hfill
    
    \includegraphics[width=0.32\textwidth]{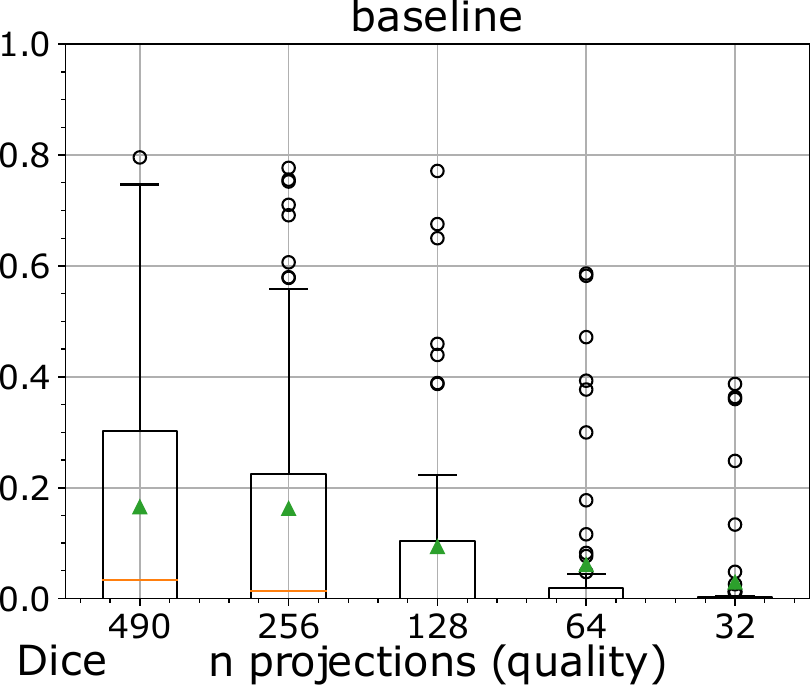}
    \includegraphics[width=0.32\textwidth]{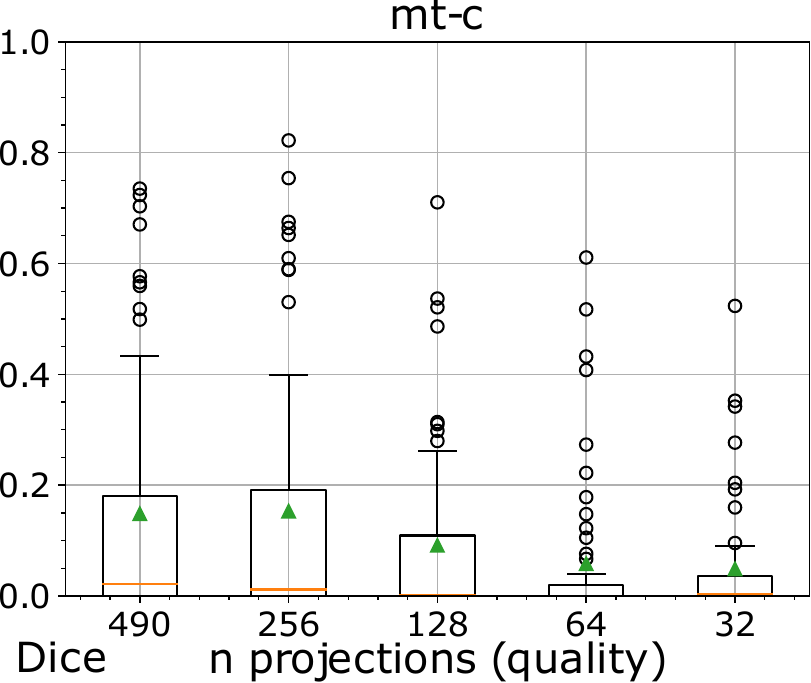}
    \includegraphics[width=0.32\textwidth]{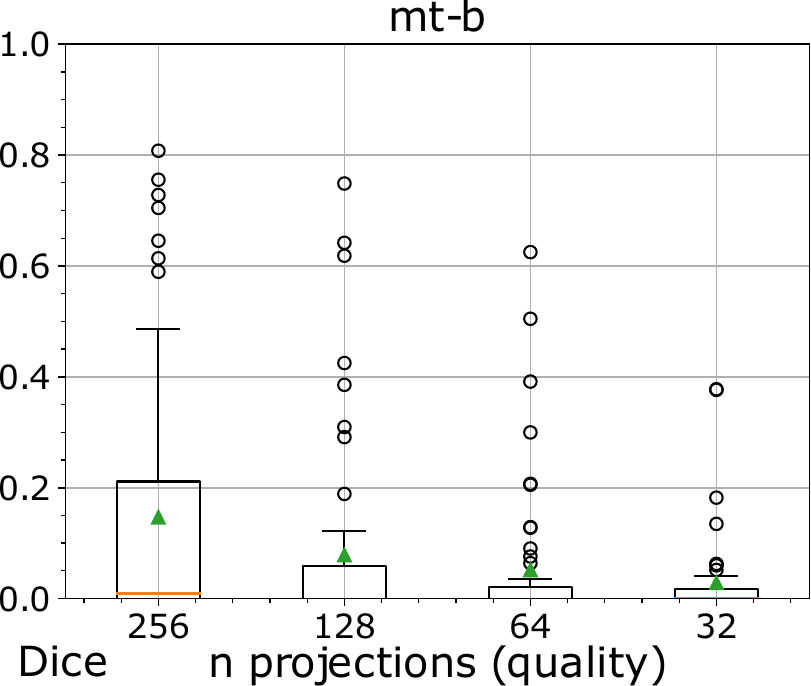} \hfill
    \caption{Holistic segmentation results using different quality levels based on the amount of projections (boxplot x-axis). The $3$ upper boxplots show the results of baseline (nn-unet), mt-c and mt-b. The top row shows Dice scores based on liver segmentation. The bottom row shows liver tumor Dice scores. The orange lines show the median and green triangles the mean.}
    \label{fig:holisticresults}
\end{figure}

\begin{figure}[ht]
    \centering
    \includegraphics[width=0.32\textwidth]{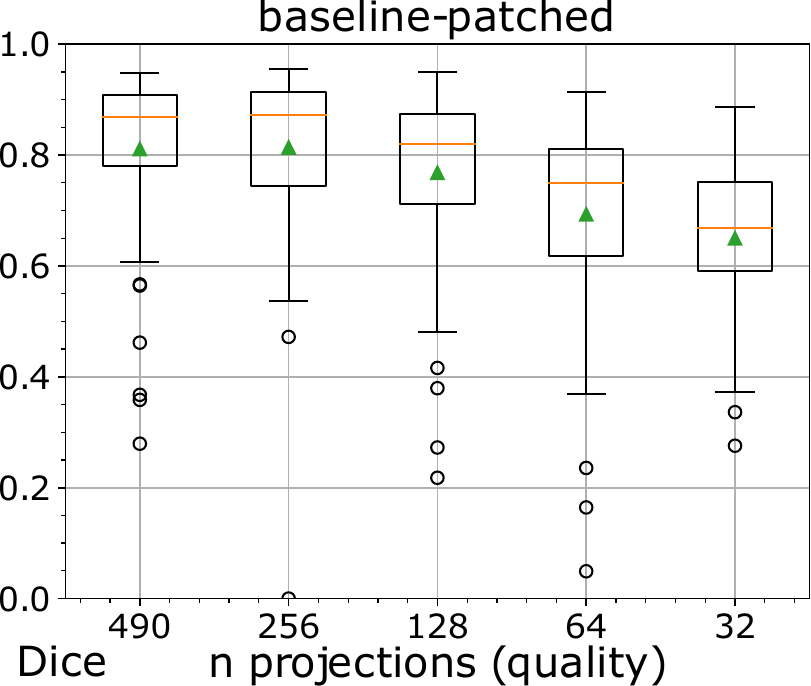}
    \includegraphics[width=0.32\textwidth]{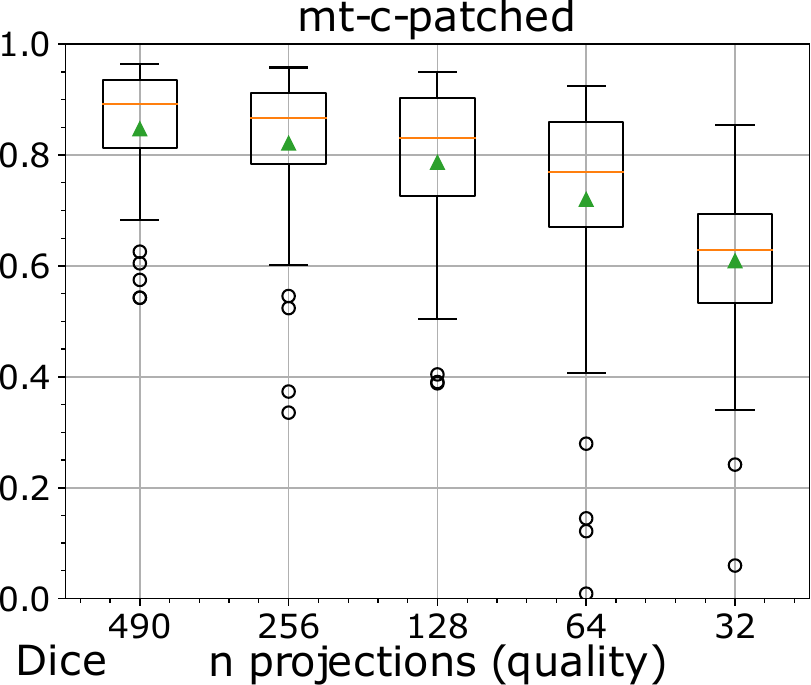}
    \includegraphics[width=0.32\textwidth]{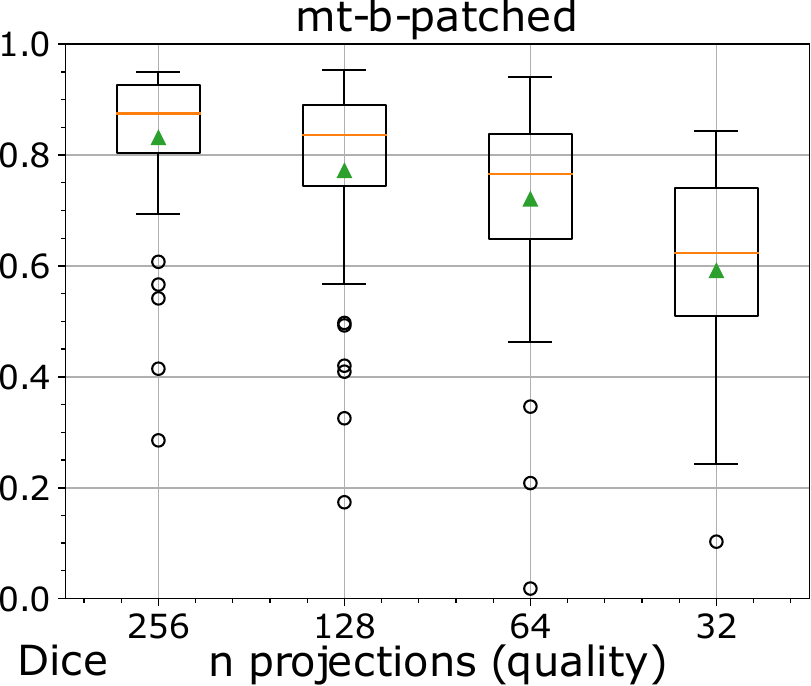} \hfill
    
    \includegraphics[width=0.32\textwidth]{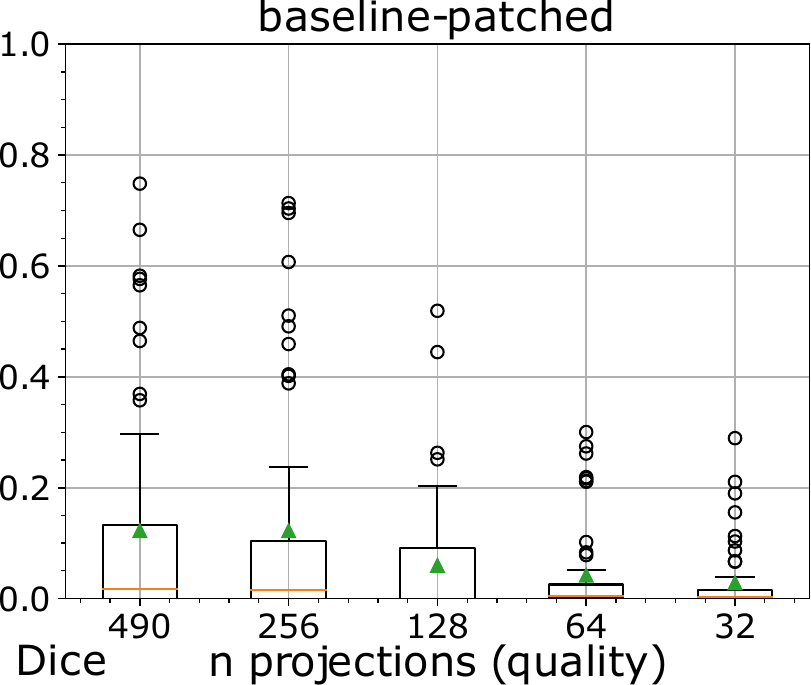}
    \includegraphics[width=0.32\textwidth]{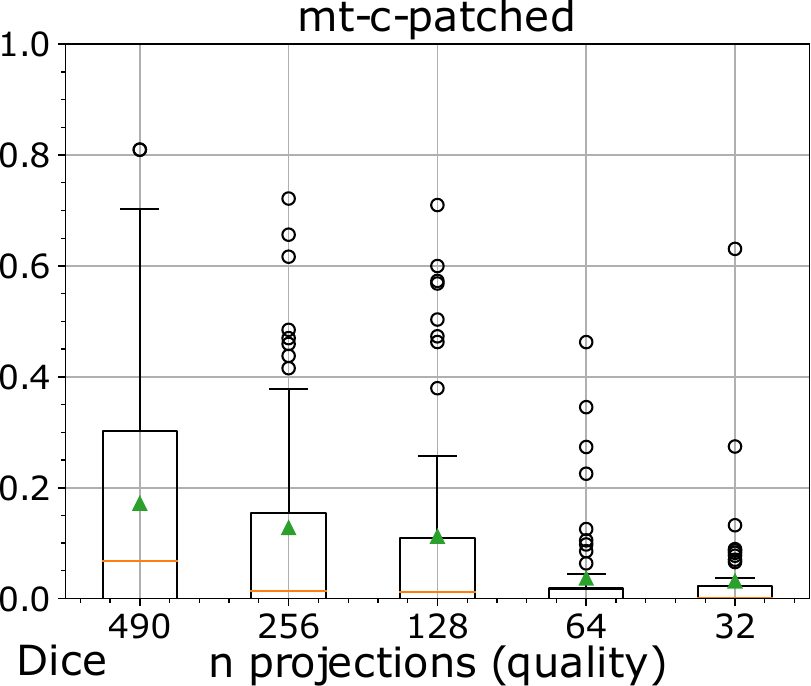}
    \includegraphics[width=0.32\textwidth]{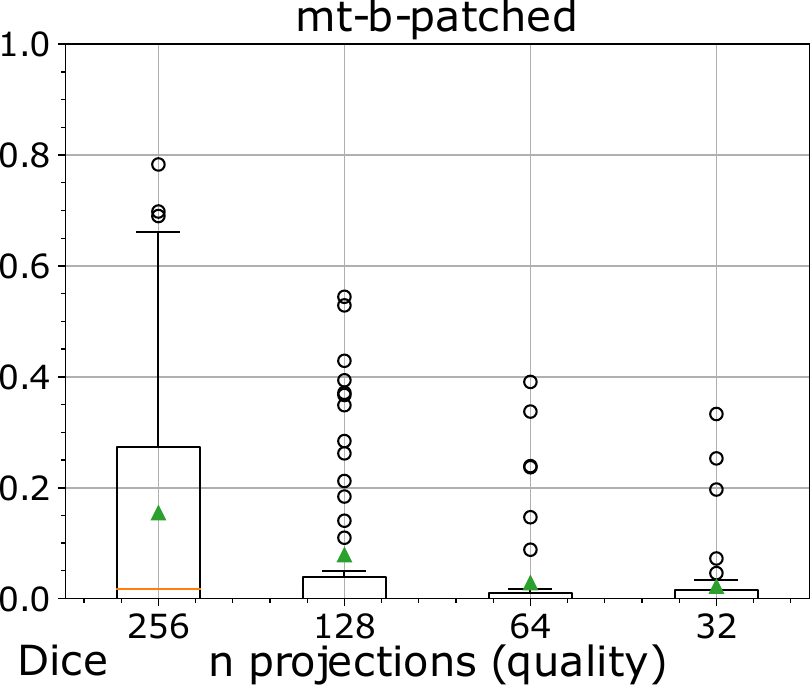} \hfill
    \caption{Patch-based segmentation results using different quality levels based on the amount of projections (boxplot x-axis). The $3$ upper boxplots show the results of our baseline (nn-unet), mt-c and mt-b. The top row shows Dice scores based on liver segmentation. The bottom row shows liver tumor Dice scores. The orange lines show the median and green triangles the mean.}
    
    \label{fig:patchedresults}
\end{figure}

The segmentation results of the holistic approach are shown in Fig.~\ref{fig:holisticresults}. It shows boxplots of segmentation results using different quality levels based on the amount of projections. The $3$ upper boxplots show the results of our baseline (nn-unet), mt-c and mt-b with the highest quality reconstruction target ($n_p=490$). 
The top row shows Dice results based on liver segmentation. The bottom row shows liver tumor Dice. Fig.~\ref{fig:patchedresults} follows the same structure for patch based liver and liver tumor segmentation.
Holistic segmentation was most successful using the baseline nn-unet with the best scores for liver segmentation with a Dice of $0.88$ for $490$ and $256$ projections and $0.78$ for $32$ projections. The mt-c approach lead to a minor improvement of $0.01$ in the case of $128$ projections and mt-b lead to an improvement of $0.02$ from $0.82$ to $0.84$ with $64$ projections. Regarding liver tumor segmentation, the baseline nn-unet performed best in all cases except $32$ projections, where mt-c proposed regularisation lead to an improvement of $0.02$.
For patch based segmentation, mt-c and mt-b show dominance. mt-c displays best values for $490$ and $128$ projections with an increase in performance to the baseline by $0.04$ and $0.02$ respectively. For $256$ and $64$ projections, mt-b led to best Dice scores with an increase of $0.02$ and $0.03$ over the baseline. The effect regarding liver tumor segmentation is similarly positive.

\section{Discussion}
Our investigation of multi-task learning for image segmentation, shows a notable distinction between holistic approaches, characterized by holistic volume understanding, and patched approaches, emphasizing high-frequency information with smaller receptive fields (in terms of real tissue). The holistic approaches generally show superior results, but the impact of multi-task learning is more pronounced in our patched approaches. We assume that the increased effect of mt-c in the patched approaches is due to the retention of high-frequency information (compared to holistic approaches). This high frequency information proves valuable for segmenting complex structures such as liver tumors. Additionally, mt-b further improves patched, full-scale approaches, suggesting that the denoising effects are more useful when high-frequency information is preserved. These findings lead to two key insights. Firstly, the results affirm that multi-task learning improves segmentation accuracy. Adding an image reconstruction tasks, especially if there is important morphological, high-frequency information, enhances the model's ability to segment complex structures. Secondly, the downscaling needed for holistic approaches might hinder the model's capacity to leverage multi-task learning fully. To further validate these findings, we propose a future investigation of full-resolution, holistic approaches to remove the possible confounding factor of volume resolution. 

In conclusion, we evaluate multi-task learning to improve 3D semantic segmentation through the addition of image reconstruction. We evaluate multiple different volume qualities and setups and show that multi-task learning can improve segmentation performance, especially in a high resolution, patched setup.


\begin{acknowledgement}
project was partly funded by the Austrian Research Promotion Agency (FFG) under the bridge project "CIRCUIT: Towards Comprehensive CBCT Imaging Pipelines for Real-time Acquisition, Analysis, Interaction and Visualization" (CIRCUIT), no. 41545455 and by the county of Salzburg under the project AIBIA.
\end{acknowledgement}

\printbibliography

\end{document}